\documentclass[conference]{IEEEtran}
\IEEEoverridecommandlockouts
\usepackage{amsmath,amssymb,amsfonts}
\usepackage{algorithmic}
\usepackage[caption=false,font=normalsize,labelfont=sf,textfont=sf]{subfig}
\usepackage{stfloats}
\usepackage{graphicx}
\usepackage{textcomp}
\usepackage{xcolor}
\usepackage[backend=bibtex,doi=false,url=false]{biblatex}
\begin{document}

\title{Implementation of a Misalignment-Tolerant MIMO Near Field Wireless Power Transfer System\\
}

\author{\IEEEauthorblockN{Taroh Hijikata\IEEEauthorrefmark{1}, Allan Mesa Jr.\IEEEauthorrefmark{2}, Charleston Dale Ambatali\IEEEauthorrefmark{3}}
\IEEEauthorblockA{\textit{Electrical and Electronics Engineering Institute} \\
\textit{University of the Philippines Diliman}\\
Quezon City, Philippines \\
\IEEEauthorrefmark{1}taroh.hijikata@eee.upd.edu.ph, \IEEEauthorrefmark{2}allan.jr.mesa@eee.upd.edu.ph, \IEEEauthorrefmark{3}charleston.ambatali@eee.upd.edu.ph}
}

\maketitle

\begin{abstract}
% Wireless power transfer (WPT) efficiency of reactive near-field systems degrade quickly with respect to distance and are also sensitive to misalignment. This limits the movement of the device being powered, thus, limiting the application of near-field WPT to non-contact cases such as electric vehicle charging. To address this issue, a multiple input multiple output (MIMO) implementation is considered due to its ability to shape the magnetic field between the transmitting and receiving coils. Maximum efficiency can be achieved by setting the correct amplitude and phase of each input coil, but achieving this setting requires a precise measurement of the S-parameters of the system. In this paper, we present an the use of the Nelder-Mead iterative optimization algorithm to estimate the maximum efficiency input setting of a 4-element transmitter 2-element receiver near-field WPT system. Using the measured S-parameters of the implementation, the proposed system is observed to significantly improve the WPT efficiency in aligned and misaligned cases.
%Applying the algorithm in the 30 cm misalignment case, efficiency improved from 1.51\% to 74.22\% at 10 cm and from 17.70\% to 72.40\% at 15 cm. A hardware setup was fabricated to experimentally verify the innate misalignment tolerance of MIMO MRC systems. At no misalignment, the efficiency is 92.51\% at 10 cm and 90.63\% at 15 cm. The efficiency at 30 cm of y-direction misalignment is 69.58\% at a distance of 10 cm and 64.52\% at a distance of 15 cm.

The efficiency of reactive near-field wireless power transfer (WPT) systems degrades rapidly with increasing separation distance and is highly sensitive to misalignment between transmitting and receiving coils. These limitations restrict the mobility of powered devices and confine many near-field WPT applications to static scenarios. To address these challenges, a multiple-input multiple-output (MIMO) WPT configuration is investigated due to its capability to shape the magnetic field distribution between the transmitter and receiver. Maximum power transfer efficiency can be achieved by appropriately setting the amplitude and phase of each transmitting coil; however, determining these optimal settings requires accurate knowledge of the system’s S-parameters. This paper presents the use of the Nelder-Mead iterative optimization algorithm to estimate the input amplitude and phase settings that maximize transfer efficiency in a near-field WPT system. The implementation comprises a four-element transmitter and a two-element receiver. Based on measured S-parameters, the proposed approach significantly improves WPT efficiency under both aligned and misaligned conditions.

\end{abstract}


\section{Introduction}

% Wireless power transfer (WPT) provides a method to migrate energy in a cordless manner. This has various applications in portable electronic devices, implanted medical devices, integrated circuits, space-based solar power satellites, electric vehicles, unmanned aerial vehicles, and smart home applications. Many devices in the market are battery-powered and these batteries have suffered from low capacity and high cost in the past \cite{wpt_overview}. Although the costs and performance of batteries have improved over the years, the increasing demand made rechargeable batteries still suffer from the same issues today. WPT provides a solution by constantly providing power to the battery even when the body to which it is connected is moving. This eases the low capacity issues of the battery and curbs the need to buy multiple batteries to compensate for the low capacity, which contributes to its high cost and high weight. WPT also helps in scenarios where traditional wired power transmission is not possible, such as long-distance charging and power transfer involving satellites.

High-efficiency wireless power transfer (WPT) is achieved in the reactive or radiated near-field zones \cite{shinohara2014wpt, near_field_wpt}. Most commercial WPT products developed over the past decade use electromagnetic induction between two coils to transfer power over the air, typically to recharge smartphones that support wireless charging. However, for single-input single-output (SISO) inductive power transfer (IPT) systems, efficiency rapidly decreases as the distance between the transmitter and receiver increases \cite{dn_synthesis}. 

In addition to distance-related issues, SISO IPT systems also suffer from misalignment sensitivity, causing rapid changes to the impedance seen by the power source and thereby reducing efficiency. This limits the operating conditions of the system, necessitating static relative positions between the source and receiver.

To address misalignment and distance issues, multiple coils can be used in different ways. One method is to use relay coils with mutual inductances tuned to accommodate a wide range of impedances \cite{peng2024optimizing}. Another is to use multiple coils as power sources \cite{zparam_efficiency}, also called a multiple-input multiple-output (MIMO) IPT system. Techniques considered in long-range radiative WPT can be applied to MIMO-IPT. The maximum theoretical efficiency can also be determined using the system’s S-parameters \cite{sparam_efficiency}. 

To determine the beam setting that achieves this (i.e., amplitude and phase setting of each coil), complete knowledge of the S-parameters is required for any given configuration. Due to the dynamic nature of contactless IPT, the computational complexity hinders the implementation of MIMO-IPT.

By treating the power transferred between coils as beams, dynamic beamforming solutions from radiative WPT can be implemented, known as dynamic beam focusing \cite{hajimiri2021dynamicbeam}. Previous work used a set of arbitrary focal points where the receiver was placed \cite{near_field_wpt}.  Others tracked receivers using directional radiation algorithms \cite{direction_radiation} or  retrodirective antennas \cite{bsrdaa_model, bsrdaa_hardware, koo2021retroreflectivearray, kato2023emimo}. 

In this paper, we present a particular MIMO-IPT implementation that utilizes the Nelder-Mead iterative optimization algorithm to determine the maximally efficient transmitter setting. The receiver coil reports its received power back to the transmitter coils, which then adjust amplitude and phase settings determined through the Nelder-Mead algorithm \cite{nelder_mead}. This process iterates until the received power converges. This removes dependency on estimating S-parameters, enabling practical use of MIMO-IPT. Furthermore, we show that MIMO-IPT is tolerant to misalignment, making it a promising candidate for contactless IPT.

\section{MIMO-IPT Implementation}

\subsection{Coil Design and Board Fabrication}

Assuming a constant Q-factor, transmission efficiency will reach 80\% if the air gap is smaller than the coil radius \cite{MRC_WPT}. Since the minimum operating distance of this work is set to 10 cm, the maximum diameter of the coil to be used is set to 20 cm.
% As MRC does not rely on strong magnetic coupling, using square coils did not result in a large decrease in efficiency. In \cite{squarejust}, square coils have a slightly higher coupling coefficient and higher maximum power capability than circular coils with the same parameters. Reference \cite{squarevspenta} also shows that square coils and circular coils with similar surface areas have similar coupling coefficients and efficiency, with square coils being slightly higher. In \cite{squarevshex}, square coils have greater tolerance against horizontal misalignment when compared to circular coils. Thus, a square layout was chosen for this project.
\figurename \ref{fig:psc_diagram} shows the geometric parameters of a square PSC. The parameters chosen for this implementation are number of turns n = 2, $s$ and $w$ = 1 cm, and $d_i$ = 14 cm. %Using the following equation from \cite{L_formula}, the computed inductance is 1.355 $\mu$H.

%\begin{equation}
%L=\dfrac{\mu_on^2d_{avg}(1.27)}{2}(ln(\dfrac{2.07}{\rho}) + 0.18\rho + 0.13\rho^2)
%\label{eq:current_sheet_approx}
%\end{equation}

\begin{figure} [t]
     \centering
     \includegraphics[width=0.2\textwidth]{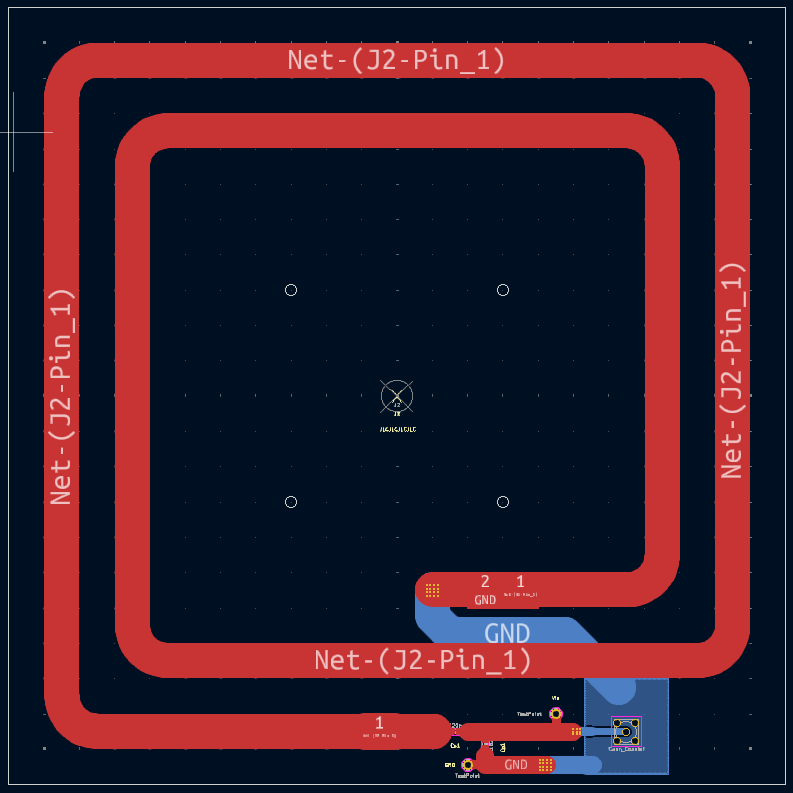}
     \caption{The PSC in KiCAD 8.0}
     \label{fig:1_setup_hardware}
\end{figure}

%The PSC to be fabricated was designed using KiCAD 8.0 and was fabricated by JLCPCB.
The fabricated PSC used FR4 as a substrate, with a board thickness of 1.6 mm, board dimensions of 24 cm x 24 cm, copper thickness $t_c$ of 35 $\mu$m, and has 2 copper layers. An SMA port was used to serve as the input for the board, and test points were placed near the traces in order for an oscilloscope probe to capture the voltage of the input for the transmitter coils, and voltage at the load for the receiver coils. The capacitors used for the impedance matching networks are surface mount device multilayer ceramic capacitors. The GERBER file of the PSC is shown in \figurename \ref{fig:1_setup_hardware}.

% \begin{figure} [t]
%     \centering
%     \includegraphics[width=0.3\textwidth]{figures/hardware psc.jpg}
%     \caption{Hardware PSC}
%     \label{fig:psc_hardware}
% \end{figure}

\subsection{Efficiency Prediction using S-parameters}

Generally, given an $M$ transmitter $\times$ $N$ receiver system, the relationship between the incident power wave and reflected power wave at all ports is given by an $(M+N) \times (M+N)$ S-parameter matrix. To calculate the maximum possible efficiency \cite{sparam_efficiency}, the S-parameters of the system are used as expressed in equation \eqref{eq:s-parameters}.

\begin{equation}
\begin{bmatrix}
    \mathbf{B}_t \\
    \mathbf{B}_r
\end{bmatrix}
=
\begin{bmatrix}
    \mathbf{S}_{tt} & \mathbf{S}_{tr} \\
    \mathbf{S}_{rt} & \mathbf{S}_{rr}
\end{bmatrix}
\begin{bmatrix}
    \mathbf{A}_t \\
    \mathbf{A}_r
\end{bmatrix}
\label{eq:s-parameters}
\end{equation}

In the above equation, $\mathbf{S}_{tt}\in\mathbb{C}^{M\times M}$ is the scattering matrix among the transmitters, $\mathbf{S}_{tr}\in\mathbb{C}^{M\times N}$ is the transfer matrix from receivers to transmitters, $\mathbf{S}_{rt}\in\mathbb{C}^{N\times M}$ is the transfer matrix from transmitters to receivers, and $\mathbf{S}_{rr}\in\mathbb{C}^{N\times N}$ is the scattering matrix among the receivers. $\mathbf{A}_t\in\mathbb{C}^{M\times 1}$ and $\mathbf{B}_t\in\mathbb{C}^{M\times 1}$ are the incident and reflected waves at the transmitters, respectively. $\mathbf{A}_r\in\mathbb{C}^{N\times 1}$ and $\mathbf{B}_r\in\mathbb{C}^{N\times 1}$ are the incident and reflected waves at the receivers, respectively. From equation \eqref{eq:s-parameters}, the maximum possible efficiency is determined using equation \eqref{eq:S_efficiency}, where $()^T$ means matrix transpose, $()^*$ means complex conjugate, and $()^H$ means complex conjugate transpose, $\mathbf{A} = \begin{bmatrix} \mathbf{A}_t & \mathbf{A}_r\end{bmatrix}^T$ describes the incident waves on the coils, and $\mathbf{E}_t$ and $\mathbf{E}_r$ are the $M \times M$ and the $N \times N$ identity matrices, respectively.

% \begin{equation}
%     P_{in} = 0.5(||A_t||^2 - ||B_t||^2)
%     \label{eq:input_power}
% \end{equation}
% \begin{equation}
%     P_{out} = 0.5(||B_r||^2 - ||A_r||^2)
%     \label{eq:output_power}
% \end{equation}

\begin{equation}
    \begin{split}
        \eta = -\dfrac{\mathbf{A}^H\mathbf{CA}}{\mathbf{A}^H\mathbf{DA}} \\
        \mathbf{C} =
    \begin{bmatrix}
        \mathbf{S}_{tr}^*\mathbf{S}_{rt} & \mathbf{S}_{tr}^*\mathbf{S}_{rr} \\
        \mathbf{S}_{rr}^*\mathbf{S}_{rt} & \mathbf{S}_{rr}^*\mathbf{S}_{rr} - \mathbf{E}_r
    \end{bmatrix} \\
    D =
    \begin{bmatrix}
        \mathbf{S}_{tt}^*\mathbf{S}_{tt} - \mathbf{S}_t & \mathbf{S}_{tt}^*\mathbf{S}_{tr} \\
        \mathbf{S}_{rt}^*\mathbf{S}_{tt} & \mathbf{S}_{rt}^*\mathbf{S}_{tr}
    \end{bmatrix}
    \end{split}
    \label{eq:S_efficiency}
\end{equation}

The expression for $\eta$ above is the generalized Rayleigh quotient, with its maximum value derived from the generalized eigenvalue equation in\eqref{eq:S_rayleigh}, where $\gamma$ are the $M+N$ eigenvalues and the columns of $\mathbf{X}$ are the $M+N$ eigenvectors. The largest eigenvalue $\gamma$ that is less than 1 is the maximum possible efficiency of the system. This serves as the baseline for evaluating the performance of the proposed iterative optimization setup.

%$\eta_{max}$ is the largest eigenvalue of $\gamma$ whose value is less than or equal to one. Defining a new matrix $B$ which is equal to $\sqrt{D}$.

\begin{equation}
    \mathbf{CX} = \gamma \mathbf{DX}
    \label{eq:S_rayleigh}
\end{equation}

%  Defining another matrix 

% \begin{equation}
%     E = B^{-H}CB^{-1}
%     \label{eq:S_efficiency_max}
% \end{equation}

% The largest eigenvalue of $E$ that is less than 1 is equal to $\eta_{max}$.

\subsection{Simulation and Implementation Setup}
CST Studio Suite 2021 was used to simulate the S-parameters of the four-transmitter, two-receiver system. L-matching networks were employed to resonate all coils at 13.56 MHz. The goal of the impedance matching networks was to reduce self-reflection of all coils in the MIMO system at the operating frequency, thereby increasing overall efficiency.

\begin{figure} [b]
    \centering
    \subfloat[]{\includegraphics[width=0.2025\textwidth]{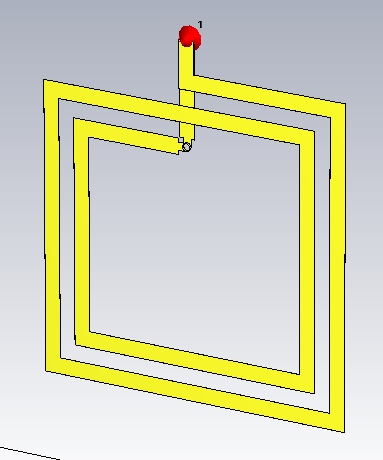}\label{fig:1_setup}}\hspace{0.3cm}
    \subfloat[]{\includegraphics[width=0.225\textwidth]{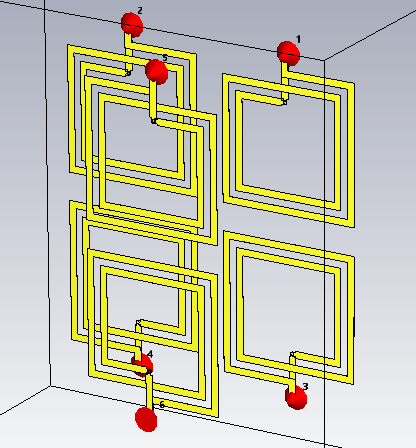}\label{fig:4x2_setup}}
    \caption{(a) The 1 coil model and (b) the 4x2 coil model implemented in CST.}
\end{figure}

The receivers were positioned 15 cm away and aligned with the transmitters. The values used for the L-matching networks are 120 pF series capacitors and 820 pF shunt capacitors. \figurename \ref{fig:1_setup} shows the single coil setup in CST while \figurename \ref{fig:4x2_setup} shows the MIMO setup.

The system's S-parameter matrix is exported to MATLAB for efficiency calculations. The maximum achievable efficiency was computed using equations \eqref{eq:S_efficiency} and \eqref{eq:S_rayleigh}. Simulations were carried out for different combinations of lateral misalignment and distance between transmitter and receiver coils. From the S-parameters, efficiency could be calculated for arbitrary settings.

\begin{figure}
    \centering
    \includegraphics[width=\linewidth]{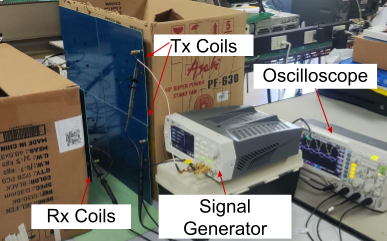}
    \caption{Hardware measurement setup.}
    \label{fig:hardware setup}
\end{figure}

\begin{figure} [t]
    \centering
    \includegraphics[width=0.45\textwidth]{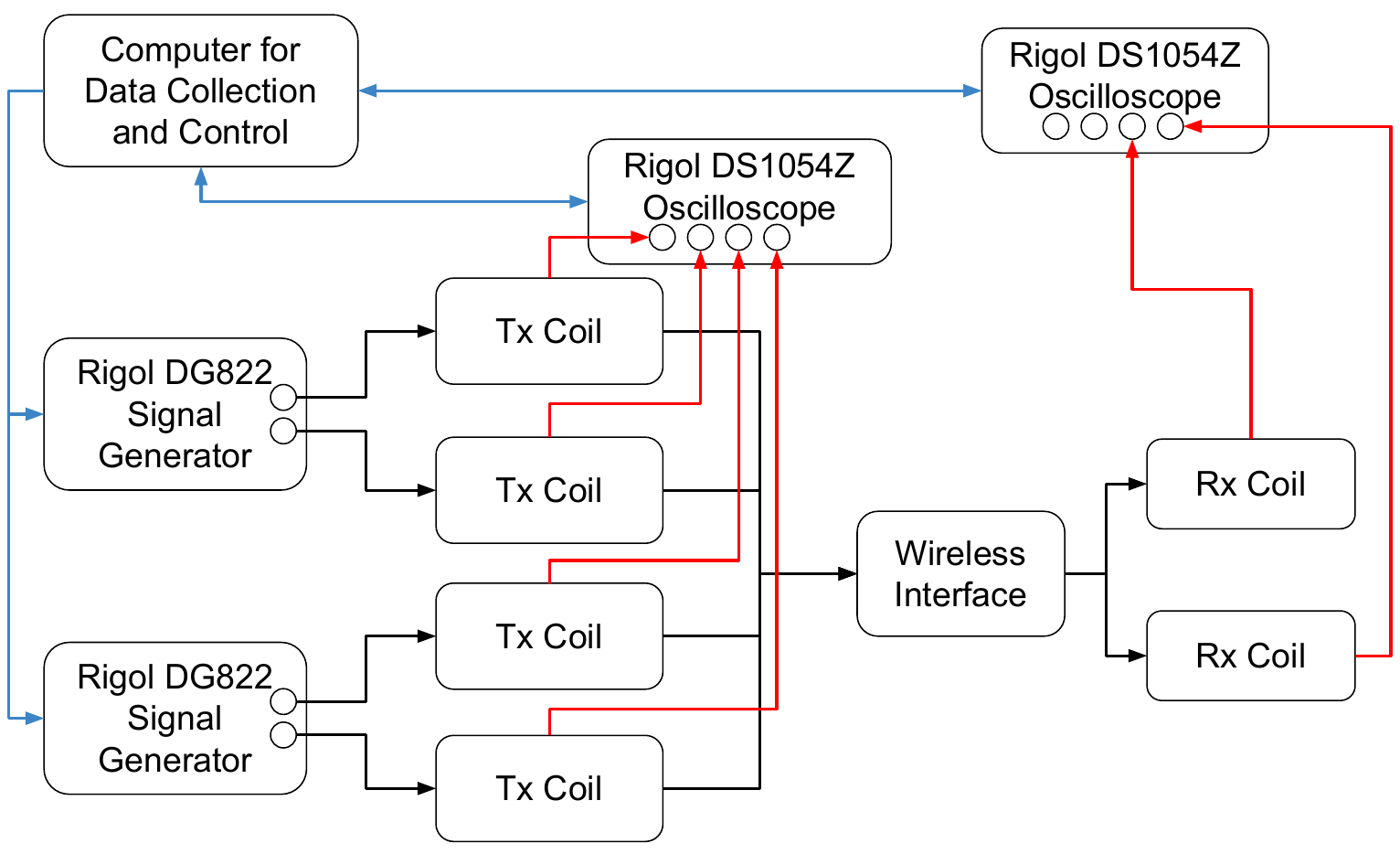}
    \caption{Block diagram of the hardware}
    \label{fig:setup block diagram}
\end{figure}

The hardware implementation is shown in \figurename \ref{fig:hardware setup} with block diagram in \figurename \ref{fig:setup block diagram}. Signal generators and oscilloscopes were controlled via a computer through a USB interface using MATLAB. 50 $\Omega$ terminators served as the dummy load for the receivers.

Oscilloscopes captured waveforms, which were stored on the computer. FFT analysis was used to extract amplitude and phase information, forming complex numbers representing the total voltage of transmitters and receivers. The amplitude and phase set on the signal generators were used to construct the incident voltage of the transmitter. The incident voltage of the receiver was set to zero, so its total voltage equaled the reflected voltage. The reflected voltage of the transmitter was computed  using $B_{t} = T_{t} - A_{t}$, where $T_{t}$ is the total voltage of the transmitter.

\subsection{Nelder-Mead Algorithm}

Starting from an arbitrary initial setup denoted by $\mathbf{A}$, the Nelder-Mead algorithm \cite{nelder_mead} transforms the initial value of $A$ into a column vector that yields good efficiency whenever there is misalignment. Details of the algorithm is available in its original paper \cite{nelder_mead}. In summary, the algorithm begins with a set of $n + 1$ vertices $x_0, x_1,..., x_n$, where $n$ is the dimension, resulting in an initial working simplex $S$. Only the incident wave of the transmitter is optimized in $A$, resulting in $n = 8$. 

The objective function $f(x)$ is evaluated at each vertex and arranged from lowest to highest. Since the goal is to maximize efficiency, the objective function used is the negative of equation (\ref{eq:S_efficiency}). Centroid $c$ is calculated using equation \eqref{eq:centroid}, assuming $f(0) \leq f(1) \leq ... \leq f(n)$.

\begin{figure} [t]
    \centering
    \includegraphics[width=0.45\textwidth]{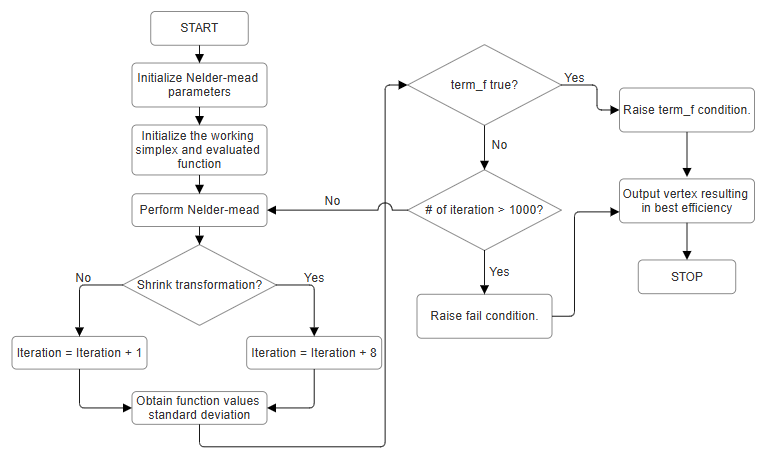}
    \caption{Procedure for the algorithm}
    \label{fig:procedure}
\end{figure}

\begin{equation}
    c = \dfrac{1}{n} \sum_{n=0}^{n-1} x_{n}
    \label{eq:centroid}
\end{equation}

The worst-performing vertex is replaced depending on the location of $c$ resulting in a new working simplex. The four transformations of the algorithm are reflection, expansion, contraction, and shrinking, relative to the best side. The reflection parameter $\alpha$ is set to 1, contraction parameter $\beta$ is set to 0.5, expansion parameter $\gamma$ is set to 2, and shrinking parameter $\delta$ is set to 0.5. If the smallest value of $f(x)$ converges, the iterations end. Else, the algorithm restarts again. This is summarized by the flowchart in \figurename \ref{fig:procedure}.

%To finish the algorithm, termination conditions need to be specified. Two conditions were used in this project. $term_f$, which becomes true when the standard deviation of the objective function is less than or equal to 0.1\%. And $fail$, which becomes true when the number of iterations exceeds 1000. Regardless of the termination condition, the vertex that results in the best efficiency will be used as the transmitter input.

%\section{Efficiency Calculations for MIMO Systems}

\section{Results and Discussions}

For both the simulated and hardware systems, data were obtained at distances of 10 cm and 15 cm. The simulation considered lateral misalignment in both the x-direction and y-direction, while the hardware setup considered misalignment only in the y-direction. In \figurename \ref{fig:4x2_setup}, the positive x-direction is to the right, and the positive y-direction is upward. Seven misalignment data points are considered from -30 cm to 30 cm. The input of all transmitters is set to an amplitude of 0.5 V and a phase of 0$^{\circ}$ for both the pre-Nelder-Mead setup and hardware. The Nelder-Mead algorithm is applied only to the simulated system and not in the hardware. 

It must be noted that the power transfer efficiency being considered in this project is the beam efficiency, not the overall efficiency of the system.

Before applying Nelder-Mead, the simulated system at a distance of 10 cm achieved a maximum efficiency of 94.24\%, at no misalignment. The lowest efficiency of the system in the y-direction is 76.61\% and the lowest efficiency in the x-direction is 1.51\%. After applying Nelder-Mead, the lowest efficiencies have improved from 76.61\% to 93.92\% and from 1.51\% to 74.22\%.

At a distance of 15 cm and before applying Nelder-Mead, the simulated system achieved a max efficiency of 91.92\% at the case of no misalignment. The lowest efficiency in the y-direction is 78.37\%, and 17.70\% in the x-direction. After applying Nelder-Mead, efficiency of the system at all points has improved. Considering the worst points, 78.37\% has improved to 91.45\%, while 17.7\% increased to 72.40\%.

The simulated system was more tolerant to misalignment in the y-direction compared to the x-direction. Regardless of misalignment direction, efficiency was, on average, higher at a distance of 15 cm. This can be attributed to the impedance matching network being designed for a 15 cm distance with zero misalignment. The plots show that the Nelder-Mead algorithm greatly improves efficiency under large misalignment. Based on the project objectives, the simulated system is deemed misalignment-tolerant in the y-direction even without Nelder-Mead.

\begin{figure} [t]
    \centering
    \includegraphics[width=0.45\textwidth]{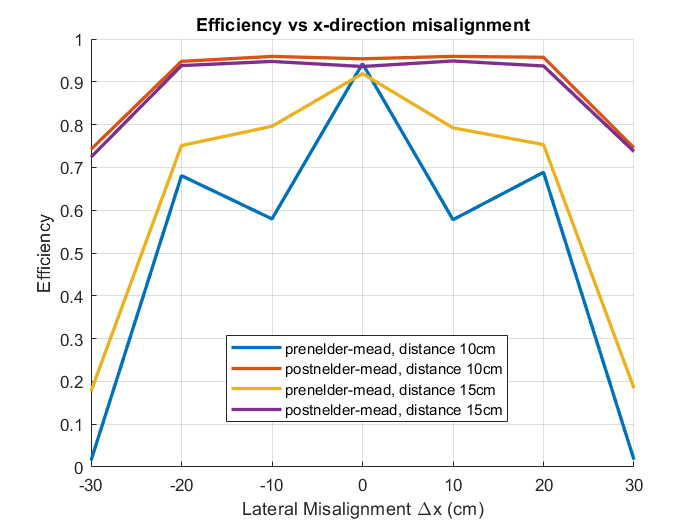}
    \caption{Efficiency of the system vs x-direction misalignment}
    \label{fig:eff_x-misalignment}
\end{figure}

\begin{figure} [t]
    \centering
    \includegraphics[width=0.45\textwidth]{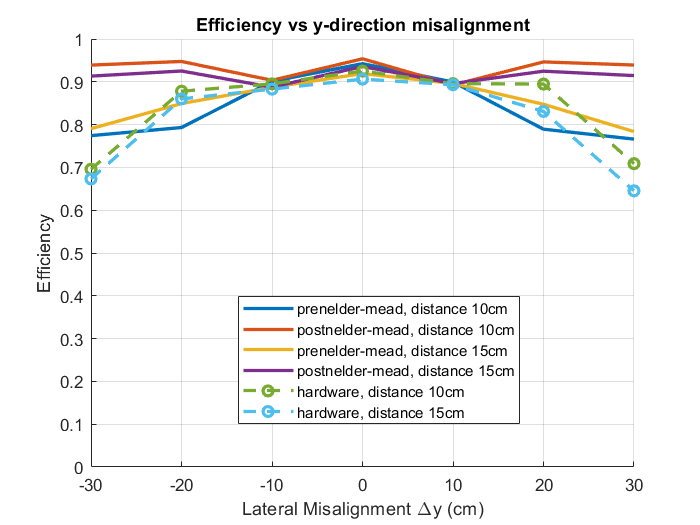}
    \caption{Efficiency of the system vs y-direction misalignment}
    \label{fig:eff_y-misalignment}
\end{figure}

At a distance of 10 cm, the hardware system achieved a max efficiency of 92.51\% at no misalignment and a minimum efficiency of 69.58\% at -30 cm of misalignment. At a distance of 15 cm, the max efficiency is 90.63\%, and the minimum efficiency is 64.52\% at 30 cm of misalignment. The hardware results also seem to be close to the simulation results, being slightly lower due to factors such as board losses, component losses, and instrument error. Although the input transmitter has the same amplitude and phases, from \figurename \ref{fig:eff_x-misalignment}, it can be seen that the efficiency of the hardware remained nearly flat from -20 cm to 20 cm before dipping at the maximum misalignment. %Based on the objectives of this project, the hardware system is also deemed to be misalignment tolerant. 
Shown in \figurename \ref{fig:eff_x-misalignment} is the plot of efficiency vs x-direction misalignment of the system while \figurename \ref{fig:eff_y-misalignment} shows the plot of efficiency vs y-direction misalignment, including the measured results.

\begin{figure}[t]
    \centering
    \includegraphics[width=\linewidth]{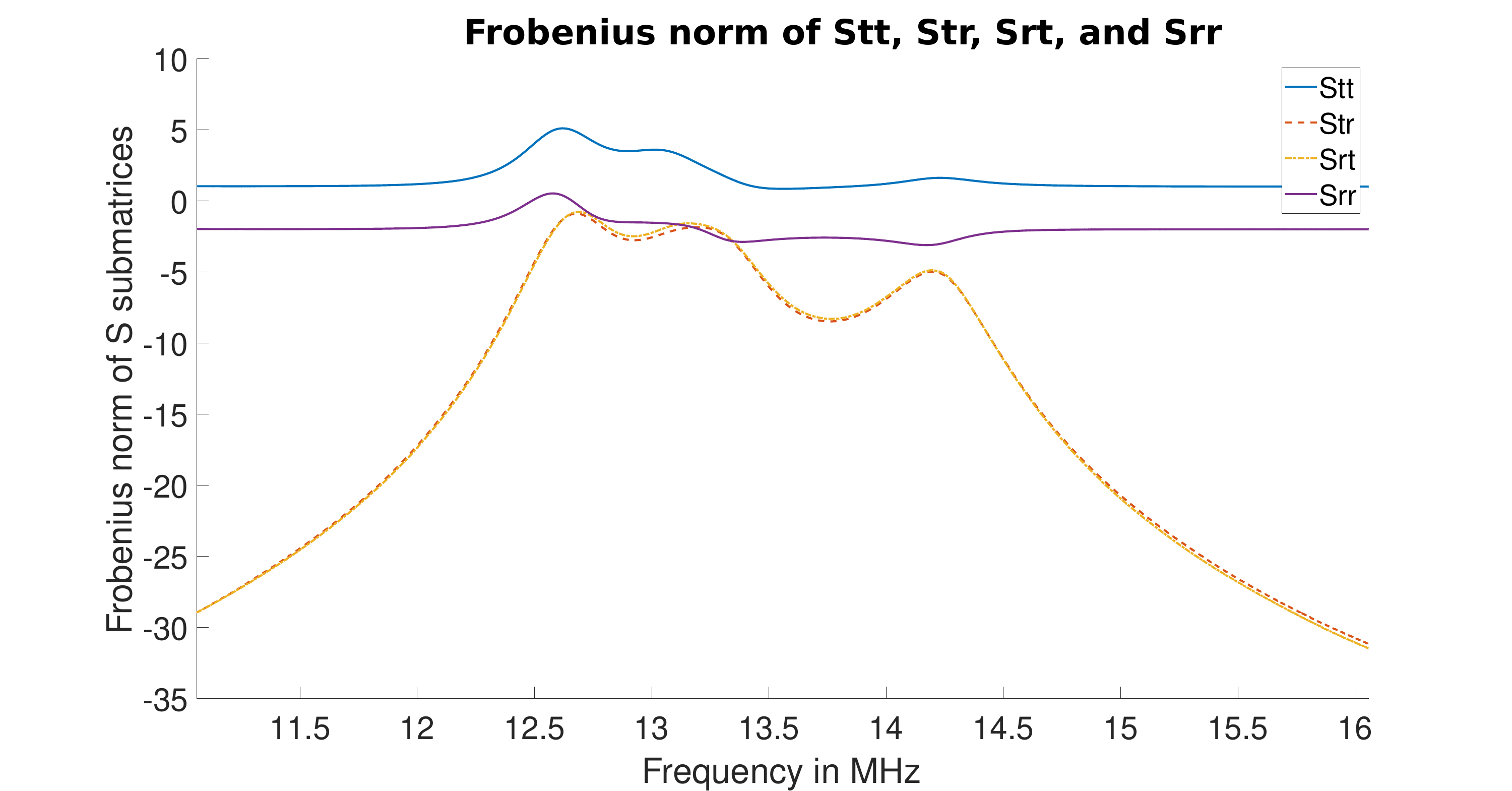}
    \caption{Frobenius norm of $\mathbf{S}$ submatrices for system with separation of 15 cm, 0 misalignment}
    \label{fig:measured_submatrices}
\end{figure}

To analyze the $S$-parameter plots of the multiport network like a two-port network, the Frobenius norms of the submatrices in Eq. \ref{eq:s-parameters} are shown in \figurename \ref{fig:measured_submatrices}. The submatrices $\mathbf{S}_{r,t}$ and $\mathbf{S}_{t,r}$ describe the coupling between the transmitter array and the receiver array. This coupling is noticeable in the frequency range of $12.5$ to $14.5~\mathrm{MHz}$. 

The submatrices $\mathbf{S}_{t,t}$ and $\mathbf{S}_{r,r}$ represent the reflection losses and mutual coupling within the transmitter and receiver arrays, respectively. These submatrices do not have a frequency range where the Frobenius norm takes a very negative value; this "dip" is expected for a well-matched antenna array. These Frobenius norms suggest that some of the energy from transmitter element goes into other transmitter elements instead of a receiver element. Although beam efficiency could be improved by applying the Nelder-Mead algorithm, mutual coupling within both transmitter and receiver arrays is still observed. 

\section{Conclusions and Recommendations}
From the results, both simulated and hardware systems are deemed misalignment-tolerant. The addition of the Nelder-Mead algorithm enhanced the simulated system’s tolerance by improving efficiency under large misalignment. Even without the Nelder-Mead algorithm, both the simulated system and hardware remained misalignment-tolerant. This is attributed to the MIMO setup and the coils’ ability to take advantage of resonance, maintaining efficient wireless power transfer.

For future work, it is highly recommended that decoupling networks \cite{dn_synthesis, mesa2025multiport} be incorporated into the transmitters of this or any other MIMO setup. Decoupling networks can be implemented either by adding additional components or by introducing another layer of coils. To increase power output, the use of class E power amplifiers and rectifiers is also recommended.

%In the case of extreme misalignment, the outer receiver coil is theorized to be receiving its power from the inner receiver coil due to cross-coupled induction. If possible, a way to separate the power measurement due to induction and due to resonance is also another recommendation.

\section*{Acknowledgment}
We acknowledge the Office of the Chancellor of the University of the Philippines Diliman, through the Office of the Vice Chancellor for Research and Development, for funding support through the PhD Incentive Award Grant 252510 YEAR 1.

\printbibliography
\clearpage

\end{document}